\documentclass[pra,aps,twocolumn,showpacs,ams]{revtex4}
\usepackage{graphicx}
\usepackage{dcolumn}
\usepackage{latexsym}
\usepackage{color}

\begin{document}


\title{
Eilenberger theory and London theory for 
transverse components \\ 
of flux line lattice form factors 
in uniaxial superconductors 
}



\author{Yuujirou Amano} 
\affiliation{
Department of Physics, Okayama University, Okayama 700-8530, JAPAN}

\author{Masahiro Ishihara} 
\affiliation{
Department of Physics, Okayama University, Okayama 700-8530, JAPAN}

\author{Masanori Ichioka}
\affiliation{
Department of Physics, Okayama University, Okayama 700-8530, JAPAN}

\author{Noriyuki Nakai} 
\affiliation{
Department of Physics, Okayama University, Okayama 700-8530, JAPAN}

\author{Kazushige Machida} 
\affiliation{
Department of Physics, Okayama University, Okayama 700-8530, JAPAN}


\date{\today}

\begin{abstract}
We theoretically study the magnetic field orientation dependence of 
longitudinal and transverse flux line lattice form factors 
in uniaxial superconductors with 
anisotropy ratio corresponding to ${\rm YBa_2Cu_3O_{7-\delta}}$. 
We discuss influences of the anisotropy ratio of coherence length, 
and differences between the $s$-wave and the $d_{x^2-y^2}$-wave pairings. 
The calculations are performed by two methods, the Eilenberger theory and 
the London theory comparatively, and we study the cutoff function of 
the extended London theory, which will be helpful in the 
analysis of the small angle neutron scattering in the vortex states. 
\end{abstract}

\pacs{74.25.Uv, 74.25.Ha, 74.72.-h, 61.05.fg}


\maketitle

\section{Introduction}

Flux line lattice (FLL) form factors in vortex states of 
type-II superconductors are observed 
by small angle neutron scattering (SANS) experiments.   
The behaviors of the form factors reflect 
exotic properties of superconductors. 
For example, 
from the temperature ($T$) dependence of the form factor　 
we can examine the existence of nodes in the superconducting gap 
function, reflecting the $T$-dependence of 
the penetration depth~\cite{Furukawa}.
From the magnetic field ($\bar{B}$) dependence, we can know the contribution 
of Pauli-paramagnetic effects in 
superconductors~\cite{DeBeer,Bianchi,IchiokaPara,White}. 
The $\bar{B}$-dependence of the FLL deformation reflects 
the anisotropy of superconductors~\cite{Suzuki}. 
In order to extract the valuable information from the SANS experiments  
in the vortex states, 
it is helpful that we perform theoretical studies 
on the behavior of the FLL form factors in the superconductors. 

In uniaxial superconductors where the coherence length is 
anisotropic between the $ab$- and the $c$-directions, 
transverse components appear in the internal fields  
when the magnetic field orientation 
is tilted from the basal plane or the $c$ axis~\cite{Thiemann}.   
While the longitudinal components $B_{z(h,k)}$ of the FLL form factor  
are obtained from the intensity of the conventional non-spin-flip SANS, 
the transverse components $B_{{\rm tr}(h,k)}$ 
are estimated from the intensity of the spin-flip SANS.  
The observation of the spin-flip SANS was performed in 
${\rm YBa_2Cu_3O_{7-\delta}}$~\cite{Kealey}  
and ${\rm Sr_2RuO_4}$~\cite{Rastovski}. 
  
Theoretically the transverse components of the FLL form factors 
were studied by the phenomenological London theory~\cite{Thiemann}.    
This method is helpful to understand the overall qualitative behaviors of 
the transverse components, 
since the FLL form factors are described by simple functions. 
However, the quantitative validity is ambiguous for    
analysis of the experimental results, 
because the vortex core contribution is neglected in the London theory.
To fix this problem, we use the extended London 
theory~\cite{Rastovski,Brandt,Yaouanc,Sonier}.   
There we introduce the cutoff function to include the vortex core 
contribution, but the detailed form and parameters of the cutoff function 
are not enough established~\cite{Yaouanc,Sonier}. 

On the other hand, by the numerical calculation based on 
the selfconsistent Eilenberger theory~\cite{Klein,IchiokaD} 
we can quantitatively 
estimate the internal fields and the FLL form factors,  
appropriately determining the vortex core structure. 
The calculation of the transverse components 
was done in the case of chiral $p$-wave 
superconductors~\cite{Ishihara,IchiokaSCES}.
The comparison of the results of the Eilenberger theory and the London theory 
was studied in the problem of the FLL morphology~\cite{Suzuki} 
and in the internal field distribution~\cite{Laiho2005,Laiho}.

\begin{figure}[b]
\begin{center}
\includegraphics[width=8.5cm]{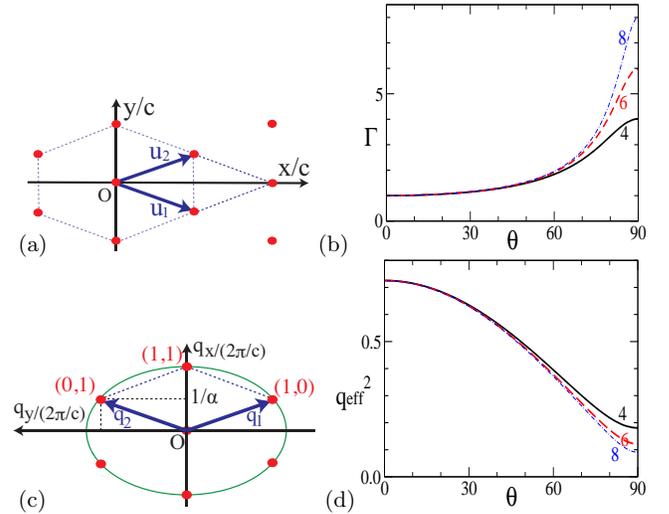}
\end{center}
\caption{\label{fig1}
(Color online) 
(a) 
FLL structure in the $(x,y)$ plane of the real space. 
Circles indicate flux line centers. 
${\bf u}_1$ and ${\bf u}_2$ are unit vectors of the FLL. 
(b) 
$\theta$-dependence of the anisotropy ratio $\Gamma_\theta$ 
for $\gamma=4$, 6, and 8.
(c) 
Circles indicate spots of the FLL form factors in the reciprocal space. 
${\bf q}_1$ and ${\bf q}_2$ are unit vectors. 
Ellipse connecting spots $(1,1)$, $(1,0)$ and the equivalent ones 
is given by $q^2_{\rm eff} =q_x^2 + (q_y/\Gamma_\theta)^2$. 
(d) 
$\theta$-dependence of $q_{\rm eff}^2$ for $\gamma=4$, 6, and 8. 
}
\end{figure}

In this paper, we investigate the magnetic field orientation dependence of 
the transverse and the longitudinal components 
of the FLL form factors for the material parameters appropriate to  
${\rm YBa_2Cu_3O_{7-\delta}}$~\cite{Kealey}. 
We perform the calculations by two methods of 
the Eilenberger theory and the London theory, comparatively.    
We study effects of the anisotropy ratio of coherence length, 
and differences between the $s$-wave and the $d_{x^2-y^2}$-wave pairings. 
To improve the agreement between results of the Eilenberger theory and 
the London theory, we discuss the cutoff functions expected in 
the extended London theory.   
In previous works~\cite{Laiho2005,Laiho} the cutoff functions were 
studied by the comparison of the internal field distribution 
between the Eilenberger theory and the London theory. 
In this work, we estimate the cutoff functions by the study on  
the magnetic field orientation dependence of the FLL form factors,   
including the transverse components.
These are directly observed in the SANS experiment.

This paper is constructed as follows. 
After the introduction, 
we explain our model for the Fermi surface and the FLL structure 
in Sec. \ref{sec:formulation}. 
The field orientation dependence of the FLL form factors is calculated 
by the Eilenberger theory in Sec. \ref{sec:Eilenberger}, 
and by the London theory in Sec. \ref{sec:London}. 
From the comparison of these results, 
we discuss the cutoff functions in the extended London theory 
in Sec. \ref{sec:ModLodndon}. 
The last section is devoted to summary. 

\section{Anisotropy ratio and flux line lattice}
\label{sec:formulation}

As a model of the Fermi surface, we use quasi-two dimensional 
Fermi surface with rippled cylinder shape,  
assuming the Fermi velocity 
${\bf v}=(v_a,v_b,v_c)\propto(\cos\phi,\sin\phi,\tilde{v}_z \sin p_c)$ 
at 
${\bf p}=(p_a,p_b,p_c)\propto(p_{\rm F}\cos\phi, p_{\rm F}\sin\phi,p_c)$
on the Fermi surface~\cite{Hiragi}.   
We consider a case $\tilde{v}_z=1/\gamma$,
to produce the anisotropy ratio of coherence length, 
$\gamma \sim  
\langle v_c^2 \rangle_{\bf p}^{1/2}/\langle v_b^2 \rangle_{\bf p}^{1/2}
\sim \xi_{c} / \xi_{b} $, 
where $\langle \cdots \rangle_{\bf p}$ 
indicates an average over the Fermi surface. 
The magnetic field orientation is tilted by $\theta$ from the $c$ axis 
toward the $b$ axis. 
Since we set the $z$ axis to be parallel to the flux lines, 
the coordinate ${\bf r}=(x,y,z)$ for the flux line structure is related to the 
crystal coordinate $(a,b,c)$  
as $(x,y,z)=(a,b \cos\theta + c \sin\theta,c \cos\theta -b \sin\theta)$.  
We set unit vectors of the FLL as 
${\bf u}_1=c({\alpha}/{2},-{\sqrt{3}}/{2},0)$ and 
${\bf u}_2=c({\alpha}/{2}, {\sqrt{3}}/{2},0)$  
with $c^2=2 \phi_0/ (\sqrt{3} \alpha \bar{B})$ and 
$\alpha=3 \Gamma_\theta$~\cite{Hiragi}, 
as shown in Fig. \ref{fig1}(a). 
$\phi_0$ is the flux quantum, and $\bar{B}$ is the flux density.  
We use the anisotropy ratio 
$\Gamma_\theta \equiv
\xi_{y} / \xi_{x} \sim 
\langle v_y^2 \rangle_{\bf p}^{1/2}/\langle v_x^2 \rangle_{\bf p}^{1/2}
\sim (\cos^2\theta+\gamma^{-2}\sin^2\theta)^{-\frac{1}{2}}$. 
Supposing the case of ${\rm YBa_2Cu_3O_{7-\delta}}$~\cite{Kealey}, 
we consider the cases of the anisotropy ratio $\gamma=4$, 6, and 8.
The $\theta$-dependence of $\Gamma_\theta$ for these cases are 
presented in Fig. \ref{fig1}(b). 
For small $\theta$, $\Gamma_\theta$ is near 1. 
On approaching $\theta \rightarrow 90^\circ$, 
$\Gamma_\theta$ rapidly increases toward $\gamma$. 

The FLL form factors  
${\bf B}({\bf q}_{(h,k)})=(B_{x(h,k)},B_{y(h,k)},B_{z(h,k)})$ 
are obtained by Fourier transformation  
of the internal field distribution ${\bf B}({\bf r})$ as  
\begin{eqnarray}
{\bf B}({\bf r})=\sum_{h,k}{\bf B}({\bf q}_{(h,k)}) 
{\rm e}^{{\rm i}{\bf q}_{(h,k)}\cdot{\bf r}}
\end{eqnarray}
with wave vector ${\bf q}_{(h,k)}=h{\bf q}_1+k{\bf q}_2$.
$h$ and $k$ are integers. 
Unit vectors in the reciprocal space 
are given by ${\bf q}_1=(2\pi/c)(1/\alpha,-1/\sqrt{3})$ 
and ${\bf q}_2=(2\pi/c)(1/\alpha,1/\sqrt{3})$. 
As presented in Fig. \ref{fig1} (c),  
the main spots $(h,k)=(1,1)$, $(1,0)$ and the equivalent ones 
are on the ellipse given by 
$q^2_{\rm eff} =q_x^2 + (q_y/\Gamma_\theta)^2$. 
At the spot $(1,1)$, $q_y=0$. 
At the spot $(1,0)$, $q_y/\Gamma_\theta=\sqrt{3}q_x$. 
The $\theta$-dependence of $q^2_{\rm eff}$ is shown in 
Fig. \ref{fig1}(d) for the cases $\gamma=4$, 6, and 8. 
 
The $z$-components $|B_{z(h,k)}|^2$ from $B_z({\bf r})$ 
give the intensity of conventional non-spin-flip SANS. 
The transverse components, 
$|B_{\rm tr(h,k)}|^2 \equiv |B_{x(h,k)}|^2+|B_{y(h,k)}|^2$,  
are accessible by spin-flip SANS experiments~\cite{Kealey,Rastovski}. 
Using the same parameters, 
we calculate the form factors by the Eilenberger theory 
and by the London theory, 
as explained in the following sections.

\section{Eilenberger theory} 
\label{sec:Eilenberger}

\begin{figure*}
\begin{center}
\includegraphics[width=15cm]{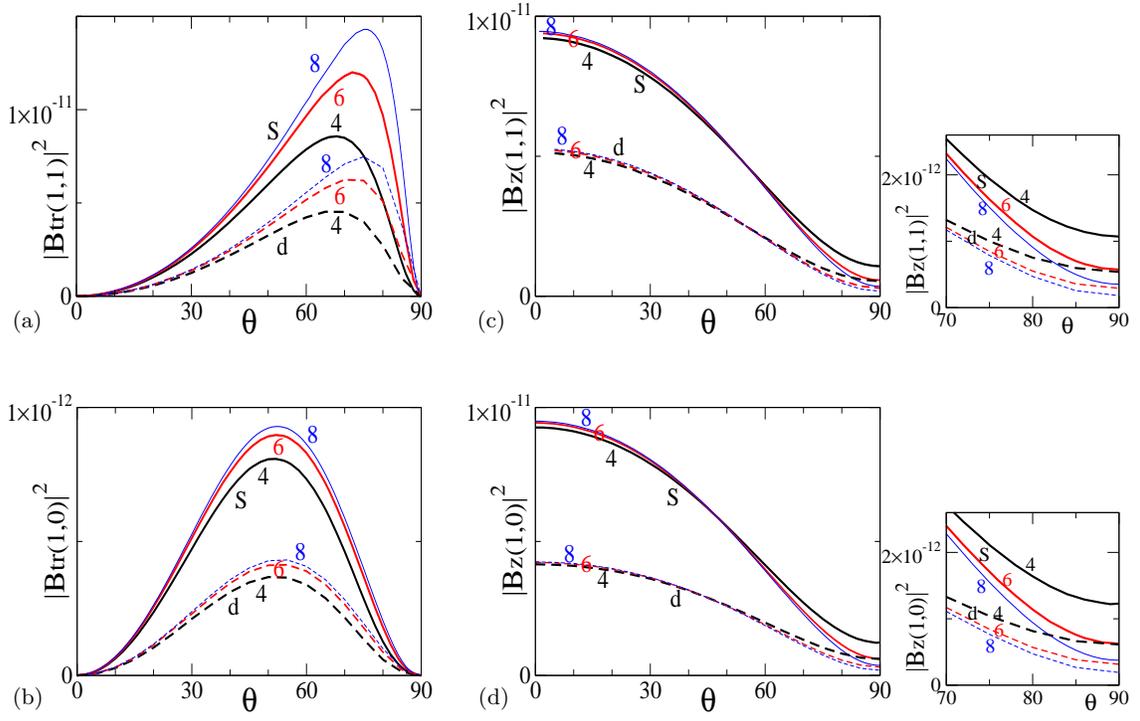}
\end{center}
\caption{\label{fig2}
(Color online) 
Field orientation dependence of the form factors by the Eilenberger theory 
for $\gamma=4$, 6, 8 in the $s$-wave pairing (solid lines) and 
the $d_{x^2-y^2}$-wave pairing (dashed lines). 
$T=0.5T_{\rm c}$ and $\bar{B}=0.1B_0$. 
As a function of $\theta$, 
we plot the transverse components 
(a) $|B_{{\rm tr}(1,1)}|^2$ for the $(1,1)$ spot, 
(b) $|B_{{\rm tr}(1,0)}|^2$ for the $(1,0)$ spot, 
and the longitudinal components 
(c) $|B_{z(1,1)}|^2$ for the $(1,1)$ spot,  
(d) $|B_{z(1,0)}|^2$ for the $(1,0)$ spot. 
In (c) and (d), 
the right panel presents focused range $70^\circ \le \theta \le 90^\circ$. 
}
\end{figure*}
\begin{figure*}
\begin{center}
\includegraphics[width=15.0cm]{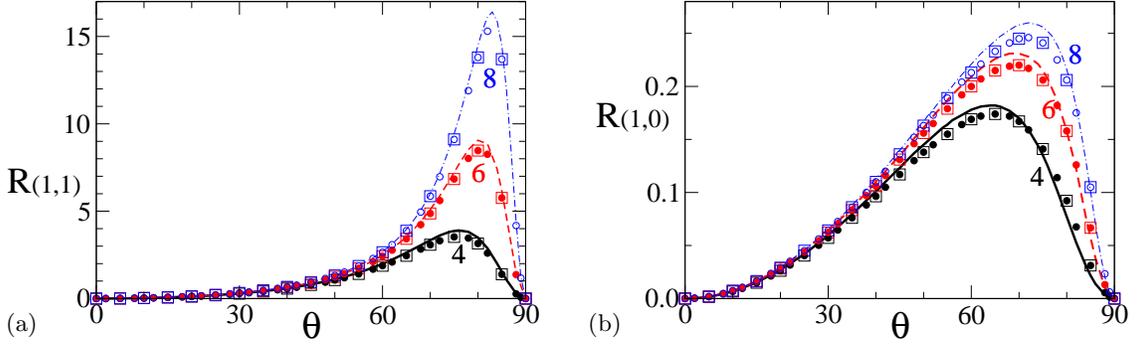}
\end{center}
\caption{\label{fig3}
(Color online)
The ratio of the spin-flip SANS intensity and the non-spin-flip one 
as a function of $\theta$ for $\gamma=4$, 6, and 8. 
(a) $R_{(1,1)}=|B_{{\rm tr}(1,1)}|^2/|B_{z(1,1)}|^2$.
(b) $R_{(1,0)}=|B_{{\rm tr}(1,0)}|^2/|B_{z(1,0)}|^2$.  
Results of the Eilenberger theory are plotted by 
circles for the $s$-wave pairing, 
and square points for the $d_{x^2-y^2}$-wave pairing. 
The lines are results by the London theory in Eq. (\ref{eq:Rhk}).
$T=0.5T_{\rm c}$ and $\bar{B}=0.1B_0$. 
}
\end{figure*}

Quasiclassical Green's functions 
$f(\omega_n , {\bf p},{\bf r})$,  
$f^\dagger(\omega_n , {\bf p},{\bf r})$,  
$g(\omega_n , {\bf p},{\bf r})$ 
are calculated in the FLL states  
by solving the Riccati equation, 
which is derived from the Eilenberger equation
\begin{eqnarray} && 
\left\{ \omega_n 
        + \hat{\bf v} \cdot 
           \left(\nabla + {\rm i}{\bf A}({\bf r}) \right) \right\} f 
= \Delta({\bf r}) \varphi({\bf p}) g  , 
\nonumber \\  && 
\left\{ \omega_n 
        - \hat{\bf v} \cdot 
           \left(\nabla - {\rm i}{\bf A}({\bf r})  \right) \right\} f^\dagger  
= \Delta^\ast({\bf r})\varphi^\ast({\bf p}) g   
\label{eq:Eil} 
\end{eqnarray} 
in the clean limit, 
with 
$ \hat{\bf v} \cdot {\nabla} g 
= \Delta^\ast({\bf r}) \varphi^\ast({\bf p}) f  
- \Delta({\bf r})\varphi({\bf p}) f^\dagger$,   
$g=(1-ff^\dagger)^{1/2}$, 
and Matsubara frequency $\omega_n$~\cite{Hiragi,Klein,Miranovic,IchiokaPara,IchiokaD,Ishihara,IchiokaSCES}. 
That is, we accurately  calculate the spatial structure of $g$ 
without using Pesch's approximation~\cite{Pesch}. 
We consider the cases of isotropic $s$-wave pairing $\varphi({\bf p})=1$
and anisotropic $d_{x^2-y^2}$-wave pairing $\varphi({\bf p})=\sqrt{2}\cos2\phi$. 
Normalized Fermi velocity is $\hat{\bf v}={\bf v}/v_{\rm F}$ with 
$v_{\rm F}=\langle {\bf v}^2 \rangle_{{\bf p}}^{1/2}$.
We have scaled length, temperature, magnetic field, 
and energies 
in units of $\xi_0$, $T_c$,
$B_0$, and $\pi k_{\rm B} T_{\rm c}$, respectively, 
where $\xi_0=\hbar v_{{\rm F}}/2\pi k_{\rm B} T_{\rm c}$ and 
$B_0=\phi_0 /2 \pi \xi_0^2$.
The vector potential 
${\bf A}({\bf r})=\frac{1}{2}\bar{{\bf B}}\times{\bf r}+{\bf a}({\bf r})$
is related to the internal field as 
${\bf B}({\bf r})=\nabla\times {\bf A}({\bf r})
 =(B_x({\bf r}),B_y({\bf r}),B_z({\bf r}))$ 
with $\bar{\bf B}=(0,0,\bar{B})$, 
$B_z({\bf r})=\bar{B}+b_z({\bf r})$ and 
$(B_x,B_y,b_z)=\nabla\times {\bf a}({\bf r})$. 
The spatial averages of $B_x$, $B_y$, and $b_z$ are zero~\cite{Rastovski}. 

The pair potential $\Delta({\bf r})$ 
is calculated by the gap equation 
\begin{eqnarray}
\Delta({\bf r})
= g_0N_0 T \sum_{0 \le \omega_n \le \omega_{\rm cut}} 
\left\langle 
 \varphi^\ast({\bf p}) \left( 
    f +{f^\dagger}^\ast 
\right) 
\right\rangle_{{\bf p}} , 
\label{eq:scD}
\end{eqnarray}
where 
$g_0$ is the pairing interaction 
in the low-energy band $|\omega_n|\le\omega_{c}$, and 
$N_0$ is the density of states at the Fermi energy in the normal state. 
$g_0$ is defined by the cutoff energy $\omega_{\rm c}$ as  
$(g_0N_0)^{-1} = \ln T+2\,T\sum_{\omega_n>0}^{\omega_{\rm c}}\,\omega_n^{-1}$.
We carry out calculations using $\omega_{\rm c}=20 k_{\rm B}T_{\rm c}$. 
Current distribution to obtain ${\bf a}({\bf r})$ is calculated by 
\begin{eqnarray}
{\bf j}_{\rm s}({\bf r})
\equiv
\nabla\times  \nabla \times {\bf a}({\bf r}) 
=-\frac{2T}{{{\kappa}}^2}  \sum_{0 \le \omega_n} 
 \left\langle \hat{\bf v}  {\rm Im}\{ g \}  
 \right\rangle_{{\bf p}} . 
\label{eq:scH} 
\end{eqnarray} 
The Ginzburg-Landau (GL) parameter 
$\kappa=B_0/\pi k_{\rm B}T_{\rm c}\sqrt{8\pi N_0}$     
is the ratio of the penetration depth $\xi_0$ 
to coherence length $\lambda_0$ for $\bar{\bf B}\parallel c$. 
In our unit $\xi_0=1$, $\kappa$ is treated as $\lambda_0$. 
In our calculations, we use $\kappa=100$ as a typical type-II superconductor.

In our study, calculations by Eqs. (\ref{eq:Eil})-(\ref{eq:scH}) 
are iterated at $T=0.5T_{\rm c}$, 
until we obtain self-consistent solutions of 
$\Delta({\bf r})$, ${\bf A}({\bf r})$, and 
quasiclassical Green's functions. 
By the selfconsistent calculations, 
we can correctly estimate the vortex core size and the core contribution 
to the internal field distribution 
${\bf B}({\bf r})=\nabla\times{\bf A}({\bf r})$.
From ${\bf B}({\bf r})$, we obtain the FLL form factors. 

In Fig. \ref{fig2}, 
we present the form factors 
as a function of the magnetic field orientation $\theta$. 
Figure \ref{fig2}(a) shows the transverse component 
$|B_{{\rm tr}(1,1)}|^2$ at the main spot $(1,1)$ in the cases of 
anisotropy $\gamma=4$, 6, 8 for the $s$-wave and 
the $d_{x^2-y^2}$-wave pairing symmetries. 
$|B_{{\rm tr}(1,1)}|^2$ has a peak at $\theta \sim 68^\circ$ ($\gamma=4$), 
$72^\circ$ ($\gamma =6$), and $76^\circ$ ($\gamma = 8$).
The transverse component reduces toward zero 
at both ends of $\theta=0$ and $\theta=90^\circ$. 
The amplitude of $|B_{{\rm tr}(1,1)}|^2$ becomes larger 
with increasing $\gamma$. 
At the other spot $(1,0)$, 
$|B_{{\rm tr}(1,0)}|^2$ in Fig. \ref{fig2}(b) is about 0.1 times smaller 
than $|B_{{\rm tr}(1,1)}|^2$ in Fig. \ref{fig2}(a), 
and the peak position shifts to angle near $\theta \sim 50^\circ$. 
The changes by an increase of $\gamma$ 
become smaller in Fig. \ref{fig2}(b). 
The longitudinal component $|B_{z(1,1)}|^2$ in Fig. \ref{fig2}(c) 
is maximum at $\theta=0$, and monotonically decreases toward 
the minimum at $\theta=90^\circ$. 
$|B_{z(1,1)}|^2$ has similar amplitude for all cases $\gamma=4$, 6, 8  
at smaller angles $\theta \le 60^\circ$, 
but it shows differences among the cases of $\gamma$    
at higher angle $60^\circ < \theta \le 90^\circ$, 
as shown in the right panel of Fig. \ref{fig2}(c). 
These behaviors are resemble to the $\theta$-dependence of 
$q_{\rm eff}^2$ in Fig. \ref{fig1}(d). 
Also at the other spot $(1,0)$, 
$|B_{z(1,0)}|^2$ in Fig. \ref{fig2}(d) has similar amplitude and 
$\theta$-dependence to those of $|B_{z(1,1)}|^2$ in Fig. \ref{fig2}(c).
In all cases of Fig. \ref{fig2}, 
the $\theta$-dependences of the $s$-wave and 
the $d_{x^2-y^2}$-wave pairing cases show similar behaviors~\cite{Kealey}, 
but the amplitudes of the $s$-wave case is about twice larger than those of 
the $d_{x^2-y^2}$-wave pairing. 

The ratio 
$R_{(h,k)} \equiv |B_{{\rm tr}(h,k)}|^2/|B_{z(h,k)}|^2$ 
is presented in Fig. \ref{fig3} by circles as a function of $\theta$ 
for spots $(h,k)=(1,1)$ and $(1,0)$. 
This corresponds to the ratio of 
the spin-flip SANS intensity to the non-spin-flip SANS intensity.  
In Fig. \ref{fig3}, the ratio in the $s$-wave pairing (circles) 
and the $d_{x^2-y^2}$-wave pairing (square points) appears on the same line. 
With increasing $\gamma$, the peak of $R_{(1,1)}$ as a function of $\theta$ 
becomes sharp, increasing the peak height rapidly. 
The peak position is $\theta=76^\circ$ ($\gamma=4$), $80^\circ$ ($\gamma=6$),  
and $83^\circ$ ($\gamma=8$). 
On the other hand, $R_{(1,0)}$ is very small compared to $R_{(1,1)}$,  
and the peak shape of $R_{(1,0)}$ is not so sharp. 
The peak positions are located at smaller $\theta$, as 
$\theta=64^\circ$ ($\gamma=4$), $70^\circ$ ($\gamma=6$), 
and $72^\circ$ ($\gamma=8$).

\section{London theory} 
\label{sec:London} 

\begin{figure*}
\begin{center}
\includegraphics[width=15cm]{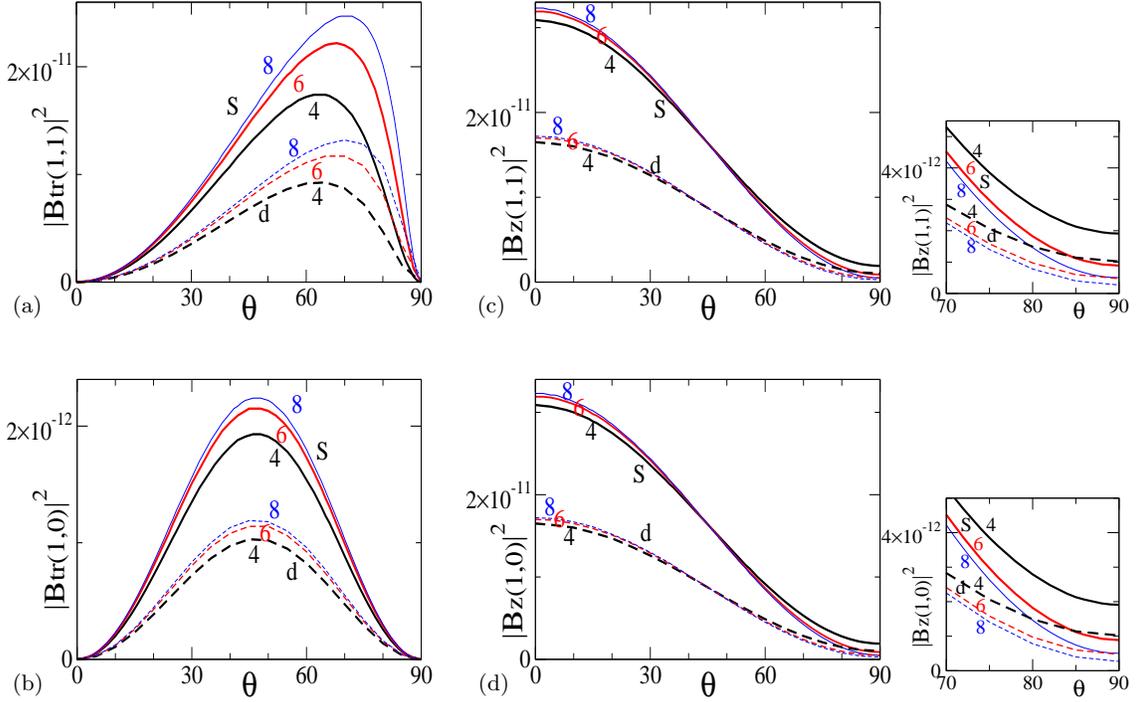}
\end{center}
\caption{\label{fig4}
(Color online) 
Field orientation dependence of the form factors by the London theory 
for $\gamma=4$, 6, 8 in the $s$-wave pairing (solid lines) and 
the $d_{x^2-y^2}$-wave pairing (dashed lines). 
$T=0.5T_{\rm c}$ and $\bar{B}=0.1B_0$. 
As a function of $\theta$, 
we  plot the transverse components 
(a) $|B_{{\rm tr}(1,1)}|^2$ for the $(1,1)$ spot, 
(b) $|B_{{\rm tr}(1,0)}|^2$ for the $(1,0)$ spot, 
and the longitudinal components 
(c) $|B_{z(1,1)}|^2$ for the $(1,1)$ spot,  
(d) $|B_{z(1,0)}|^2$ for the $(1,0)$ spot. 
In (c) and (d), 
the right panel presents focused range $70^\circ \le \theta \le 90^\circ$. 
}
\end{figure*}

The FLL form factors for uniaxial superconductors 
were studied in Ref. \onlinecite{Thiemann} 
on the basis of the London theory. 
Following the method, we evaluate the magnetic field orientation dependence 
of the form factors for the same parameters as in the previous section. 
The relation of current and vector potential 
in the reciprocal space is given as 
\begin{eqnarray}
{\bf j}_{\rm s}({\bf q})
={\rm i}{\bf q}\times{\bf B}({\bf q}) 
= - \frac{1}{\kappa^2} \hat{Q}{\bf a}({\bf q}) 
\label{eq:L01}
\end{eqnarray}
in the London theory~\cite{Suzuki,Kogan}, 
where $(i,j)$ component of the tensor $\hat{Q}$ is given by 
\begin{eqnarray}
Q_{i,j}=2T \sum_{0 \le \omega_n}
\left\langle \hat{v}_i \hat{v}_j 
\frac{|\Delta \varphi({\bf p})|^2}{\beta^3}\right\rangle_{\bf p}
\label{eq:L02}
\end{eqnarray}
with $\beta=(\omega_n^2+|\Delta \varphi({\bf p})|^2)^{1/2}$, and  
$i,j=x,y,z$. 
Here, small non-local correction terms are neglected. 
$\Delta$ is determined by the gap equation (\ref{eq:scD}) 
in the uniform state without vortices.  
In the uniaxial superconductors, 
components of the inverse matrix $\hat{m}=\hat{Q}^{-1}$ are zero 
except for $m_{xx}=m_a$, $m_{yy}=m_b \cos^2\theta +m_c \sin^2\theta$, 
$m_{yy}=m_b \sin^2\theta +m_c \cos^2\theta$, 
$m_{yz}=m_{zy}= (m_b -m_c) \sin\theta \cos\theta $, where 
\begin{eqnarray}
m_\alpha^{-1}
=2T \sum_{0 \le \omega_n}
\left\langle \hat{v}_\alpha^2 
\frac{|\Delta \varphi({\bf p})|^2}{\beta^3}\right\rangle_{\bf p}
\end{eqnarray}
for $\alpha=a,b,c$. 
$m_b^{-1}=m_a^{-1}$. 
The values of $m_a^{-1}$ and $m_c^{-1}$ are listed in Table \ref{tab:mi}.
Since $\gamma \sim (m_a^{-1}/m_c^{-1})^{1/2}$, 
$m_c^{-1}$ decreases with increasing $\gamma$. 
The $T$-dependence of $m_\alpha^{-1}$ corresponds 
to that of the superfluid density. 
In the limit $T\rightarrow 0$, 
$m_\alpha^{-1}=\left\langle \hat{v}_\alpha^2 \right\rangle_{\bf p}$, 
so that $m_\alpha^{-1}$ are independent from 
the pairing function $\varphi({\bf p})$ at $T=0$. 
At finite temperatures, 
$m_\alpha^{-1}$ in the $d_{x^2-y^2}$-wave pairing is smaller than 
that of the $s$-wave pairing, 
because the $T$-dependence of the superfluid density is different 
depending on the pairing function.  

In order to obtain ${\bf B}({\bf q})$, 
we substitute 
${\bf a}({\bf q})={\rm i}\kappa^2 \hat{m} {\bf q}\times{\bf B}({\bf q}) $ 
from Eq. (\ref{eq:L01}) to the relation 
${\bf B}({\bf q}) - {\bf q} \times {\bf a}({\bf q})=\bar{B} {\bf e}_z$, 
and use ${\bf q}\cdot{\bf B}({\bf q})=0$,  
where ${\bf e}_z$ is 
the unit vector along the $z$-direction. 
Thus, finally  ${\bf B}({\bf q})$ is written as 
\begin{eqnarray} &&
B_x({\bf q})=-\frac{\kappa^2 m_{yz} q_x q_y}{d} \bar{B} , 
\label{eq:Bx}
\\ && 
B_y({\bf q})=\frac{\kappa^2 m_{yz} q_x^2}{d} \bar{B} , 
\label{eq:By}
\\ && 
B_z({\bf q})=\frac{1+\kappa^2 m_{zz} q^2}{d} \bar{B} 
\label{eq:Bz}
\end{eqnarray} 
with 
\begin{eqnarray} && 
d=\{ 1+\kappa^2(m_{xx}q_y^2+m_{yy}q_x^2) \} (1+\kappa^2 m_{zz} q^2) 
\nonumber \\ && \qquad 
   - \kappa^4 m_{yz}^2 q^2 q_x^2 
\end{eqnarray}
and $q^2=q_x^2+q_y^2$. 
Compared to Ref. \onlinecite{Thiemann}, $x$ and $y$ axes are exchanged 
in our definition. 

The form factors by Eqs. (\ref{eq:Bx})-(\ref{eq:Bz}) are presented 
in Fig. \ref{fig4} as a function of $\theta$. 
The overall behaviors of the $\theta$-dependence by the London theory 
are resemble to those by the Eilenberger theory in Fig. \ref{fig2}. 
The amplitude of the transverse components are enhanced 
with increasing $\gamma$.  
$|B_{{\rm tr}(1,1)}|^2$ is 10 times larger than $|B_{{\rm tr}(1,0)}|^2$. 
In the longitudinal components, 
$|B_{z(1,1)}|^2$ and $|B_{z(1,0)}|^2$ are almost the same amplitude. 
Quantitative comparison of the $\theta$-dependence between 
the Eilenberger theory and the London theory is discussed 
in the next section. 
In all cases of Fig. \ref{fig4}, values of the $s$-wave pairing case 
are about twice larger than those of the $d_{x^2-y^2}$-wave pairing case. 
These dependences on the pairing symmetry in the London theory 
qualitatively accord with those in the Eilenberger theory. 
And the dependences come form the fact that 
$m_a^{-1}$ and $m_c^{-1}$ are smaller in the $d_{x^2-y^2}$-wave pairing case, 
as shown in Table \ref{tab:mi}. 

The ratio of the spin-flip SANS intensity to the non-spin-flip SANS intensity  
is given by 
\begin{eqnarray}
R_{(h,k)} \equiv \frac{|B_{{\rm tr}(h,k)}|^2}{|B_{z(h,k)}|^2}
=\frac{\kappa^4 m_{yz}^2 q^2 q_x^2}{(1+\kappa^2 m_{zz} q^2)^2}. 
\label{eq:Rhk}
\end{eqnarray}
Therefore, in the type-II limit $\kappa \gg 1$, 
\begin{eqnarray}
R_{(1,1)} =\frac{m_{yz}^2}{m_{zz}^2} 
=\left(\frac{(1-\gamma^2)\sin\theta\cos\theta}
           {\sin^2\theta +\gamma^2 \cos^2\theta}\right)^2 
\label{eq:R11}
\end{eqnarray}
at the spot $(1,1)$ where $q_y=0$, and 
\begin{eqnarray}
R_{(1,0)} =\frac{m_{yz}^2q_x^2}{m_{zz}^2q^2} 
=\frac{R_{(1,1)}}{1+3 \Gamma_\theta^2}
\label{eq:R10}
\end{eqnarray}
at the spot $(1,0)$. 
The $\theta$-dependences of $R_{(h,k)}$ in Eq. (\ref{eq:Rhk}) 
are presented by lines in Fig. \ref{fig3} for $(h,k)=(1,1)$ and (1,0). 
The $\theta$-dependences and the $\gamma$-dependences 
of $R_{(1,1)}$ and $R_{(1,0)}$ by the London theory 
(lines) give  a nice fitting to the results of the Eilenberger theory 
(circles and squares).
There we find only 
a small deviation near the peaks in the $\theta$-dependence of $R_{(1,1)}$, 
and $R_{(1,0)}$ is slightly small (large) at larger (smaller) $\theta$, 
compared to the results of the Eilenberger theory.  
The $\theta$-dependence of $R_{(1,0)}$ in Fig. \ref{fig3}(b) 
seems to correspond to 
the results reported in Ref. \onlinecite{Kealey}. 


\begin{table}
\begin{tabular} {|c|c|c|c|c|} \hline
\multicolumn{2}{|c|}{} & $T=0$ & \multicolumn{2}{c|}{$T=0.5T_{\rm c}$}
  \\ \cline{4-5}
\multicolumn{2}{|c|}{} &       & $s$-wave & $d$-wave   \\ \hline 
$\gamma=4$ & $m_a^{-1}$ 
           & $4.85\times 10^{-1}$
           & $4.03\times 10^{-1}$
           & $2.94\times 10^{-1}$ \\ 
           & $m_c^{-1}$
           & $3.01\times 10^{-2}$
           & $2.50\times 10^{-2}$ 
           & $1.83\times 10^{-2}$ \\ \hline
$\gamma=6$ & $m_a^{-1}$
           & $4.93\times 10^{-1}$
           & $4.10\times 10^{-1}$
           & $2.99\times 10^{-1}$ \\
           & $m_c^{-1}$
           & $1.37\times 10^{-2}$ 
           & $1.14\times 10^{-2}$ 
           & $8.32\times 10^{-3}$ \\ \hline
$\gamma=8$ & $m_a^{-1}$ 
           & $4.96\times 10^{-1}$
           & $4.12\times 10^{-1}$
           & $3.01\times 10^{-1}$ \\ 
           & $m_c^{-1}$
           & $7.74\times 10^{-3}$
           & $6.43\times 10^{-3}$ 
           & $4.70\times 10^{-3}$ \\ \hline
\end{tabular}
\caption{\label{tab:mi} 
$m_a^{-1}$ and $m_c^{-1}$ at $T=0$ and $0.5T_{\rm c}$ 
in the $s$-wave and the $d_{x^2-y^2}$-wave pairings for $\gamma=4$, 6, 8. 
The values at $T=0$ 
are same for the $s$-wave and the $d_{x^2-y^2}$-wave pairings. 
}
\end{table}

\begin{table}
\begin{tabular} {|c|c|c|c|} \hline
 &  \multicolumn{2}{c|}{${\displaystyle |F_{\rm tr(1,1)}|^2}$}
 &  ${\displaystyle |F_{\rm z(1,0)}|^2}$ 
  \\ \cline{2-4}
 & $s$-wave & $d$-wave & $d$-wave   \\ \hline 
(a) $\gamma=4$ & $c_1=0.861$ & $c_1=0.974$ & $c_1=0.873$ \\ 
               & $c_2=0.610$ & $c_2=0.410$ & $c_2=0.883$ \\ \hline 
(b) $\gamma=6$ & $c_1=0.831$ & $c_1=0.963$ & $c_1=0.924$ \\ 
               & $c_2=0.690$ & $c_2=0.469$ & $c_2=0.843$ \\ \hline 
(c) $\gamma=8$ & $c_1=0.807$ & $c_1=0.949$ & $c_1=0.936$ \\  
               & $c_2=0.735$ & $c_2=0.505$ & $c_2=0.836$ \\ \hline
\end{tabular}
\caption{\label{tab:fitting} 
Fitting values of parameters $c_1$ and $c_2$ in Eq. (\ref{eq:fitting}) 
for each case of Fig. \ref{fig6}. 
}
\end{table}

\begin{figure*}
\begin{center}
\includegraphics[width=11.0cm]{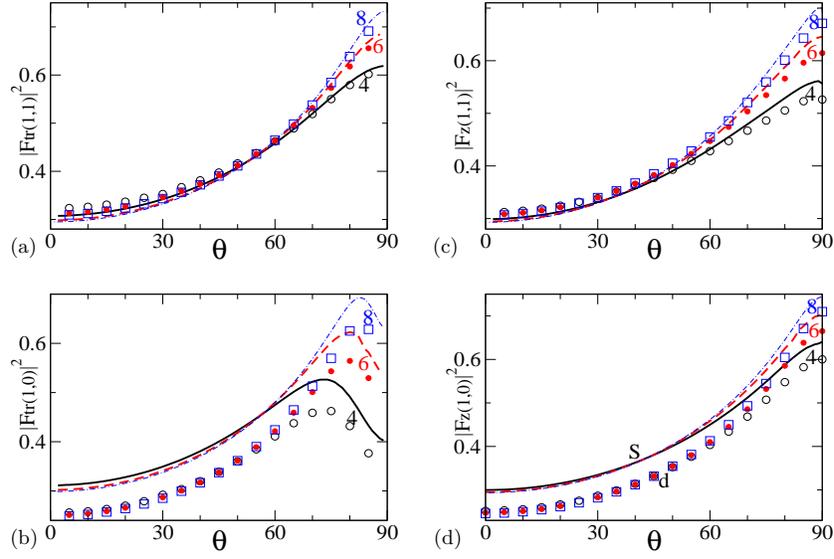}
\end{center}
\caption{\label{fig5}
(Color online) 
$\theta$-dependence of the cutoff functions defined in Eq. (\ref{eq:Cutoff}) 
for $\gamma=4$ 
(solid line), 6 (dashed line), and 8 (dashed-dotted line) 
in the $s$-wave pairing. 
We also show the $d_{x^2-y^2}$-wave pairing case for 
$\gamma=4$ ($\circ$), 6 ($\bullet$), and 8($\Box$).
(a) $|F_{{\rm tr}(1,1)}|^2$. 
(b) $|F_{{\rm tr}(1,0)}|^2$. 
(c) $|F_{z(1,1)}|^2$. 
(d) $|F_{z(1,0)}|^2$. 
$T=0.5T_{\rm c}$ and $\bar{B}=0.1B_0$. 
}
\end{figure*}
\begin{figure*}
\begin{center}
\includegraphics[width=16cm]{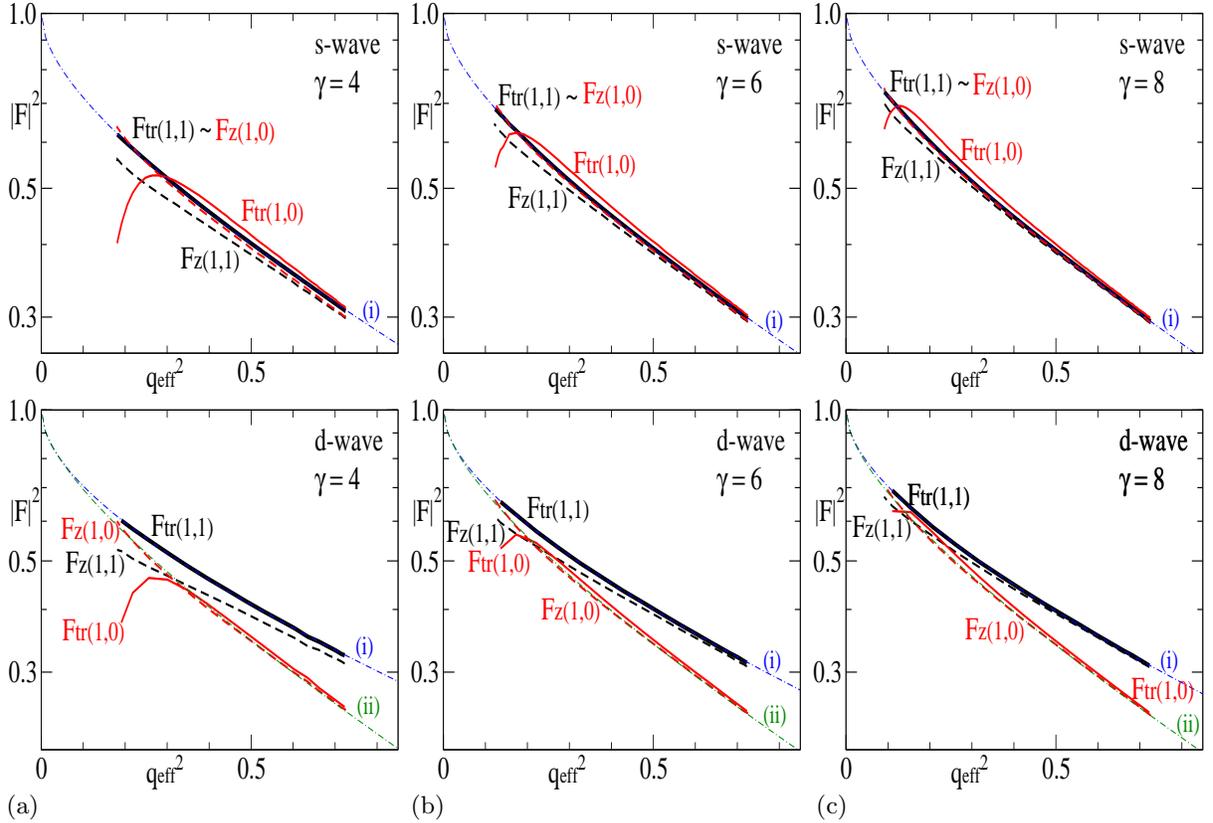}
\end{center}
\caption{\label{fig6}
(Color online) 
$q_{\rm eff}^2$-dependence of the cutoff functions  
defined in Eq. (\ref{eq:Cutoff}) 
for $\gamma=4$ (a), 6 (b), 8 (c) in the $s$-wave pairing (upper panels)
and the $d_{x^2-y^2}$-wave pairing (lower panels).
The vertical axis is a logarithmic scale. 
$T=0.5T_{\rm c}$ and $\bar{B}=0.1B_0$. 
Bold solid lines are for $|F_{{\rm tr}(1,1)}|^2$ and $|F_{{\rm tr}(1,0)}|^2$. 
Bold dashed lines are for $|F_{z(1,1)}|^2$ and $|F_{z(1,0)}|^2$. 
In the $s$-wave pairing, 
$|F_{{\rm tr}(1,1)}|^2 \sim |F_{z(1,0)}|^2$.  
The dashed-dotted lines (i) 
present fitting lines of Eq. (\ref{eq:fitting}) 
for $|F_{{\rm tr}(1,1)}|^2$.  
In the $d_{x^2-y^2}$-wave pairing, 
we also show fitting lines for $|F_{z(1,0)}|^2$ by dashed-dotted lines (ii). 
The fitting parameters are presented in Table \ref{tab:fitting}. 
}
\end{figure*}
\begin{figure*}
\begin{center}
\includegraphics[width=12cm]{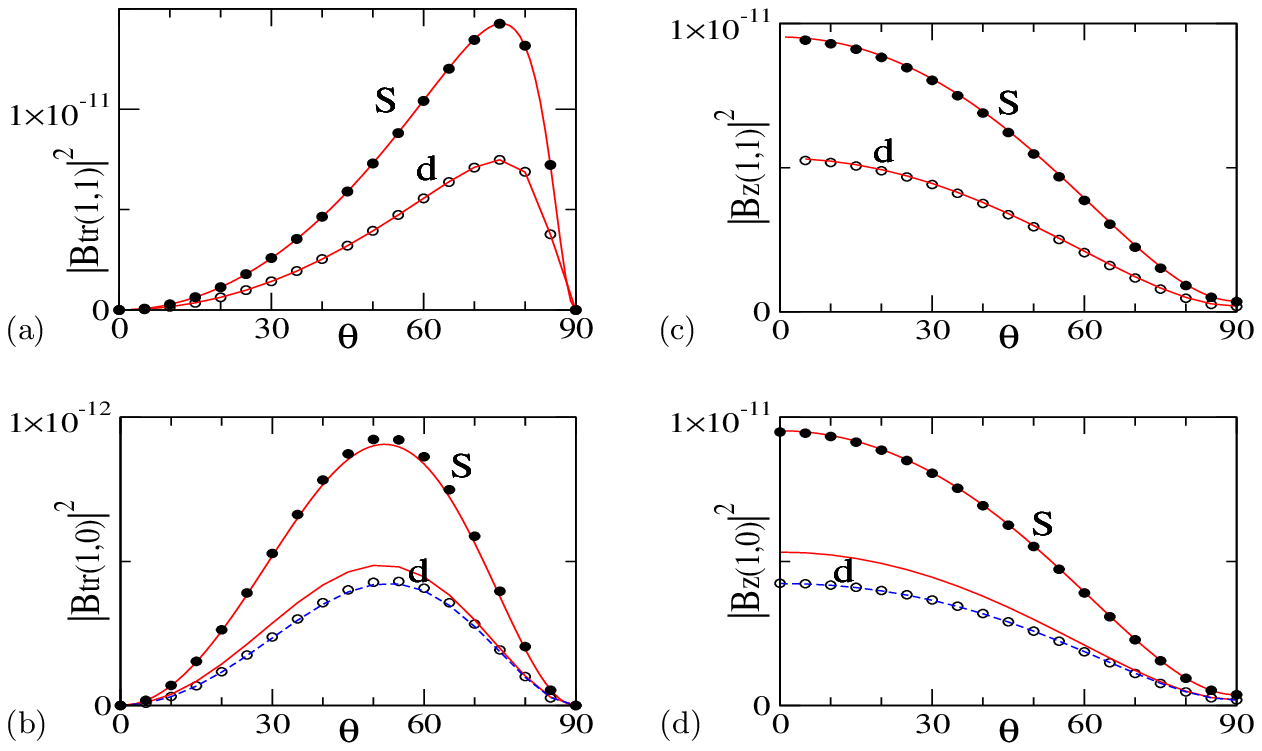}
\end{center}
\caption{\label{fig7}
(Color online) 
$\theta$-dependence of the form factors 
by the Eilenberger theory (circles) and 
the extended London theory (lines)  
for $\gamma=8$ in the $s$-wave and the $d_{x^2-y^2}$-wave pairings. 
(a) $|B_{{\rm tr}(1,1)}|^2$. 
(b) $|B_{{\rm tr}(1,0)}|^2$. 
(c) $|B_{z(1,1)}|^2$. 
(d) $|B_{z(1,0)}|^2$. 
$T=0.5T_{\rm c}$ and $\bar{B}=0.1B_0$. 
For solid lines, we use values of $c_1$ and $c_2$ for the fitting to 
$|F_{{\rm tr}(1,1)}|^2$ in Table \ref{tab:fitting}. 
For the $d_{x^2-y^2}$-wave pairing, we also show the cases of 
$c_1$ and $c_2$ for the fitting to $|F_{z(1,0)}|^2$ 
by dashed lines in (b) and (d). 
}
\end{figure*}

\section{Extended London theory}
\label{sec:ModLodndon}

In this section, we study 
a quantitative comparison of the 
form factors between the Eilenberger theory and the London theory, 
and discuss cutoff functions in the extended London theory. 
In the extended London theory, the form factors are given by 
\begin{eqnarray} && 
|B_{{\rm tr}(h,k)}|^2 = |F_{{\rm tr}(h,k)} B_{{\rm tr}(h,k)}^{(L)}|^2 , \qquad 
\nonumber
\\ && 
|B_{z(h,k)}|^2 = |F_{z(h,k)} B_{z(h,k)}^{(L)}|^2 , \qquad 
\label{eq:FFmL}
\end{eqnarray}
with introducing cutoff functions $F_{{\rm tr}(h,k)}$ and $F_{z(h,k)}$
in order to consider the vortex core contributions.  
$|B_{{\rm tr}(h,k)}^{(L)}|^2$ and $|B_{z(h,k)}^{(L)}|^2$ 
are the form factors in Fig. \ref{fig4} obtained by the London theory. 

Since the form factors calculated by the Eilenberger theory are quantitatively 
reliable, we substitute them to 
$B_{z(h,k)}$ and $B_{{\rm tr}(h,k)}$ in Eq. (\ref{eq:FFmL}). 
Thus, we estimate the cutoff functions as 
\begin{eqnarray} && 
|F_{{\rm tr}(h,k)}|^2= |B_{{\rm tr}(h,k)}^{(E)}|^2 / |B_{{\rm tr}(h,k)}^{(L)}|^2 , 
\nonumber 
\\ && 
|F_{z(h,k)}|^2= |B_{z(h,k)}^{(E)}|^2 / |B_{z(h,k)}^{(L)}|^2 , 
\label{eq:Cutoff}
\end{eqnarray}
where 
$|B_{{\rm tr}(h,k)}^{(E)}|^2$ and $|B_{z(h,k)}^{(E)}|^2$ 
are the form factors in Fig. \ref{fig2} obtained by the Eilenberger theory. 

Figure \ref{fig5} presents the $\theta$-dependence of 
the cutoff functions $|F_{{\rm tr}(h,k)}|^2$ and $|F_{z(h,k)}|^2$ for 
the spots $(h,k)=(1,1)$ and $(1,0)$. 
These are increasing functions as a function of $\theta$, 
except for $|F_{{\rm tr}(1,0)}|^2$ near $\theta=90^\circ$. 
The changes by an increase of $\gamma$ appear at $\theta > 60^\circ$, 
where the cutoff functions become larger for larger $\gamma$. 
At the spot $(1,1)$, the $s$-wave and the $d_{x^2-y^2}$-wave pairing cases 
have similar values for $|F_{{\rm tr}(1,1)}|^2$ (a) and $|F_{z(1,1)}|^2$ (c). 
At the spot $(1,0)$, the $d_{x^2-y^2}$-wave pairing cases 
have smaller values for $|F_{{\rm tr}(1,0)}|^2$ (b) and $|F_{z(1,0)}|^2$ (d), 
compared to the $s$-wave pairing case. 

We assume the cutoff functions in the form 
\begin{eqnarray}  && 
|F_{{\rm tr}(h,k)}|^2, 
|F_{z(h,k)}|^2 = {\rm exp}(-c_1 q_{\rm eff} -c_2 q_{\rm eff}^2). 
\label{eq:fitting}
\end{eqnarray} 
For the fitting, we plot $|F_{{\rm tr}(h,k)}|^2$ and $|F_{z(h,k)}|^2$ 
for $(h,k)=(1,1)$ and $(1,0)$ 
as a function of $q_{\rm eff}^2$ in Fig. \ref{fig6}, 
where the vertical axis is a logarithmic scale. 
Usually the Gaussian form ($c_1=0$) is used as a  
conventional cutoff function~\cite{Rastovski,Brandt,Yaouanc,Sonier}. 
If the Gaussian function is used, 
the fitting lines in Fig. \ref{fig6} become straight lines.  
However, we include a term $-c_1 q_{\rm eff}$ in the exponent 
to satisfy the condition 
$\lim_{q_{\rm eff}\rightarrow 0} |F_{z(h,k)}|^2 =1$,  
so that 
$\lim_{q_{\rm eff}\rightarrow 0} B_{z(h,k)} =B_{z(0,0)}=\bar{B}$.
As shown in Fig. \ref{fig6}, 
$|F_{{\rm tr}(1,1)}|^2$ is well fitted by the function 
in Eq. (\ref{eq:fitting}). 
The fitting parameters $c_1$ and $c_2$ for each panel of Fig. \ref{fig6} 
are listed in Table \ref{tab:fitting}. 
In the $s$-wave pairing, 
$|F_{z(1,0)}|^2 \sim |F_{{\rm tr}(1,1)}|^2$.  
While $|F_{z(1,1)}|^2$ and $|F_{{\rm tr}(1,0)}|^2$ are 
also near the fitting line, 
they show small deviations for smaller $q_{\rm eff}^2$. 
In the $d_{x^2-y^2}$-wave pairing, 
we also show the fitting curve for $|F_{z(1,0)}|^2$, 
since the fitting functions for the $(1,1)$ spot 
and for the $(1,0)$ spot have different slopes in Fig. \ref{fig6}. 
This indicates that the anisotropy ratio of the vortex core shape 
is slightly deviated from $\Gamma_\theta$ in this range of $T$ and $\bar{B}$, 
due to the node structure of the $d_{x^2-y^2}$-wave pairing. 
That is, the dependence of the fitting function is changed from 
$q_{\rm eff}^2=q_x^2 +(q_y/\Gamma_\theta)^2$ to 
$q_x^2 +c_y (q_y/\Gamma_\theta)^2$. 
The factor $c_y( \ne 1)$ reflects the different cutoff of the vortex core size 
between the $x$- and the $y$-directions. 
Since $q_y=0$ at the $(1,1)$ spot, only the spot $(1,0)$ includes 
the influence of $c_y$. 
Also the deviations of $|F_{{\rm tr}(1,0)}|^2$ at smaller $q_{\rm eff}^2$ 
indicate that $|F_{{\rm tr}(1,0)}|^2$ depends on the variable such as 
$q_x^2 +c_y (q_y/\Gamma_\theta)^2$, 
and the factor $c_y$ has the $\theta$-dependence near $\theta=90^\circ$  
to cancel the rapid change of $\Gamma_\theta$.  

Finally, for the comparison to the results by the Eilenberger theory, 
in Fig. \ref{fig7} we show the $\theta$-dependence of the form factors 
of Eqs. (\ref{eq:FFmL}) and (\ref{eq:fitting}) by the extended London theory 
for $\gamma=8$. 
There we use the fitting parameters in Table \ref{tab:fitting}. 
In the $s$-wave pairing, all of four form factors in Fig. \ref{fig7} 
by the Eilenberger theory are well fitted by the extended London theory 
using the same fitting parameters $c_1$ and $c_2$. 
In the $d_{x^2-y^2}$-wave pairing, we also see the nice fitting 
by the extended London theory, 
while we have to change parameters $c_1$ and $c_2$ 
for the $(1,0)$ spot (dashed lines) 
from those for the $(1,1)$-spot (solid lines). 
These fittings suggest the importance of vortex core contribution 
in the estimate of the cutoff function in the extended London theory. 

\section{Summary}
\label{sec:summary}

In summary, 
we studied the magnetic field orientation dependence 
of the transverse and longitudinal FLL form factors, 
and clarified changes by anisotropy ratio of uniaxial superconductors.
We also evaluated contributions of the pairing symmetry, 
considering $s$-wave and $d_{x^2-y^2}$-wave pairings. 
The $d_{x^2-y^2}$-wave pairing case has smaller form factors than those of 
the $s$-wave pairing case, reflecting the different $T$-dependence 
of the superfluid density. 
These evaluations were performed by two methods; 
Eilenberger theory and London theory. 
The former is quantitatively reliable, and the latter is a simple formulation. 
Comparing results of two theories, we found that cutoff function is 
necessary to modify the London theory for quantitative analysis 
of spin-flip and non-spin-flip SANS experiments. 

The cutoff function reflects the contribution of the vortex core 
in the internal magnetic field. 
As future studies, we have to estimate the cutoff functions  
at other $\bar{B}$ and $T$ ranges, 
and examine the $\bar{B}$- and $T$-dependences of 
the vortex core contributions. 
Also in the SANS experiments, if the experimental data of the form factors 
are substituted in Eq. (\ref{eq:FFmL}), 
the behaviors of the cutoff functions are evaluated experimentally. 
From these future studies, we hope to clarify the vortex core contributions 
in the longitudinal and transverse internal field distributions, 
including the dependences on the pairing symmetry.

\begin{acknowledgments}
We thank M. R. Eskildsen for fruitful discussions and 
information about spin-flip SANS experiments.
\end{acknowledgments}


\end{document}